\documentclass[12pt]{article}

\usepackage[dvips]{graphicx}

\newcommand{\beq}{\begin{eqnarray}}
 \newcommand{\eeq}{\end{eqnarray}}
\newcommand{\be}{\begin{equation}}
 \newcommand{\ee}{\end{equation}}

\newcommand{{\SD}}{\rm SD}

\newcommand{\vex}{\mbox{\boldmath${\rm x}$}}
\newcommand{\vew}{\mbox{\boldmath${\rm w}$}}
\newcommand{\vey}{\mbox{\boldmath${\rm y}$}}
\newcommand{\ver}{\mbox{\boldmath${\rm r}$}}

\newcommand{\vez}{\mbox{\boldmath${\rm z}$}}

\newcommand{\lan}{\langle}
\newcommand{\ran}{\rangle}

\title{Quarks and baryons in QCD at finite density}
\author{ Yu.A. Simonov, M.A.Trusov\\
\textit{ITEP, Moscow, Russia}}

\date{}

\begin{document}
\maketitle

\begin{abstract}
The mechanism  of string creation for light quarks developed
earlier   is considered for nonzero  quark chemical potential. A
strong modification of the confining string  due to finite quark
density (chemical quark potential $\mu$) is observed. As a
surprising result in a multiquark system with a common string
junction an attractive well  appears of radius $\mu/\sigma$ and of
an average depth equal to $\mu$, which implies formation of
multiquark hadrons. Preliminary estimates predict a new phase
transition to multiquark hadron phase at rather high densities (in
heavy ion collisions or in neutron stars) when neutron matter is
compressed to 3-4 normal nuclear densities.


\end{abstract}

\section{Introduction}

Recently the high-density effects in QCD attracted a lot of
attention  because  of the possible density
 phase transition \cite{1,2}. On the fundamental level the physical expectation of any  phase
 transition  may  be connected to the possible reconstruction
 of the vacuum  and  for that one needs that the energy density to be
  of the order of the vacuum energy  density $\varepsilon_{cr}\sim \varepsilon_{vac} \approx
-\frac{11}{3} N_c\frac{\alpha_s}{32\pi} \lan (F^a_{\mu\nu}
 )^2\ran, \varepsilon_{cr} \sim 1$ GeV/fm$^3$ which provides the
 drastic change in the vacuum structure and may cause the phase
 transition. In case of temperature phase transition the
 corresponding energy density $\varepsilon$ is indeed of the
 order of $\varepsilon_{cr}$,  and as it was first argued
 in \cite{3}  and later
 measured on the  lattice \cite{4}, the QCD vacuum is strongly
 transformed in a way, that most part of  colorelectric fields
 evaporate above $T_c$. In this way
 $T_c$ was calculated through $\varepsilon_{cr}$ and  found in good agreement with lattice data \cite{5}, see \cite{6} for  a review.
 It was shown earlier in \cite{7}, that Chiral Symmetry Breaking (CSB) occurs
  due to confinement (more explicitly its colorelectric  scalar components),
  which means that CSB should disappear simultaneously with confinement. In another way  in the framework
  of hadron resonance gas and making use of
the low-energy theorems at $T \neq 0$ a similar  conclusion was
obtained through  application of the effective dilaton Lagrangian
to gluodynamics and to QCD \cite{8}.

  The same type of   arguments for the density phase
 transition would imply that at the baryon density of $\sim 1$
nucleon/fm$^3$, i.e. $3 \div 6$ times higher than the standard nuclear density, the vacuum can be reconstructed in such a way, that part of
fields, e.g. the colorelectric fields responsible for confinement, disappear above critical density. The actual calculations in \cite{5} have
supported this conclusion and give critical chemical potential $\mu_q (T=0) =0.6$ GeV.

This is the picture of deconfinement due to high density. However,
of much more practical importance can be possible phase
transitions in the confining region,  at $2\div 3$  normal nuclear
densities.

 It is the
purpose of the present paper to start the investigation of the
role of density on the vacuum fields in general and confinement in
particular.

In this  paper we ask ourselves a short and simple question: how
the nonzero baryonic chemical potential $\mu$ acts on the
confinement of light quarks, and come  to the  unexpected answer,
that the  confining string of light quarks is destroyed gradually
by $\mu$ in such a way, that the one part of string,  near the
string junction is eaten by the nonzero $\mu $, while at distant
$r$  the string survives. We show that this can make multiquark
states (MQS) more advantageous and thus cause a new type of phase
transition. The plan of the paper is as follows. In section 2 the
effective Lagrangian is derived from the QCD Lagrangian with
nonzero $\mu$ and their solution is discussed in section 3.
Section 4 generalizes results to the case of baryons and
multiquark bags.
 Section 5 is devoted to the physical implications of results and prospectives.

\section{Derivation of Effective $4q$ Lagrangian}

  One starts with the QCD partition function  in
   presence of quark chemical potential $\mu$ in the Euclidean
  space-time, and we begin with the zero temperature, $T=0$.

  \be
Z=\int DAD\psi D\psi^+ e^{-S_0(A)+\int~^f\psi^+(i\hat
\partial+im- i\mu\gamma_4+g\hat A)~^f\psi d^4x}
\label{1}\ee where $S_0(A)=\frac{1}{4}\int(F^a_{\mu\nu}(x))^2
d^4x$, $m$ is the current quark mass (mass matrix $\hat m$ in
SU(3)), and the quark operator $^f\psi_{a\alpha}(x)$ has flavor
index $a(f=1,... n_f)$, color index $a(a=1, ... N_c)$ and Lorenz
bispinor index $\alpha(\alpha=1,2,3,4)$, and  we use the contour
gauge \cite{11} to express $A_\mu(x)$ in terms of $F_{\mu\nu}$.
 One has for the contour $z_\mu(s,x)$ starting at point $x$ and ending at
$Y=z(0,x)$
 \be A_\mu(x)=\int^1_0 ds\frac{\partial
z_\nu(s,x)}{\partial s}\frac{\partial z_\rho(s,x)}{\partial x_\mu}
F_{\nu\rho} (z(s))\equiv \int^x_Y d \Gamma_{\mu\nu\rho} (z)
F_{\nu\rho} (z) .
\label{2} \ee
 Integrating out the gluonic fields
$A_{\mu} (x)$, one obtains
\be
Z=\int D\psi D\psi^+
e^{\int~^f\psi^+(i\hat \partial+im- i\mu\gamma_4)^f\psi d^4x}
e^{L^{(2)}_{EQL}+L^{(3)}_{EQL}+...}
\label{3} \ee
 where the EQL
proportional to $\lan\lan A^n\ran\ran$ is denoted by
$L_{EQL}^{(n)}$,
\be L^{(2)}_{EQL}=\frac{g^2}{2}\int
d^4xd^4y~^f\psi^+_{a\alpha}(x)
~^f\psi_{b\beta}(x)~^g\psi^+_{c\gamma}(y)~^g\psi_{d\varepsilon}(y)
\lan  A^{(\mu)}_{ab}(x) A^{(\nu)}_{cd}(y)\ran
\gamma^{(\mu)}_{\alpha\beta} \gamma^{(\nu)}_{\gamma\varepsilon}
\label{4} \ee
Average of gluonic fields can be computed using
(\ref{2}) as (see \cite{12} for  details of derivation)
\be g^2\lan A^{(\mu)}_{ab}(x) A^{(\nu)}_{cd}(y)\ran=
\frac{\delta_{bc}\delta_{ad}}{N_c} \int^x_0
du_i\alpha_{\mu}(u)\int^y_0 dv_k\alpha_\nu(v)
D^{(E,H)}(u-v)(\delta_{\mu\nu}\delta_{ik}-\delta_{i\nu}\delta_{k\mu}),
\label{5} \ee
 where $D^{(E,H)}(x)$ is the correlator $\lan E_i(x)
E_i(0)\ran$ or $\lan H_i(x) H_i(0)\ran$. As it was argued in
\cite{12} the dominant contribution at large distances from
 the static antiquark is given by the color-electric fields,
 therefore at the first stage  we shall write down
 explicitly $L_{EQL}^{(2)} (el)$ for this case, i.e. taking $\mu=\nu=4$. As a result one
 has\cite{12}
\be
 L_{EQL}^{(2)}(el)=\frac{1}{2N_c}\int d^4x\int
d^4y~^f\psi^+_{a\alpha}(x)~^f\psi_{b\beta}(x)
~^g\psi^+_{b\gamma}(y)~^g\psi_{a\varepsilon}(y)
\gamma^{(4)}_{\alpha\beta}\gamma^{(4)}_{\gamma\varepsilon} J^E(x,y)
\label{6} \ee
where $J^E(x,y)$ is
\be
 J^E(x,y) =\int^x_0 du_i\int^y_0
dv_i D^E(u-v),~~i=1,2,3.
 \label{7} \ee
 One can form bilinears
$\Psi^{fg}_{\alpha\varepsilon}\equiv
~^f\psi^+_{a\alpha}~^g\psi_{a\varepsilon}$ and project using Fierz
procedure given isospin and Lorentz structures,
$\Psi^{fg}_{\alpha\varepsilon}\to \Psi^{(n,k)}(x,y).$ Here we
consider only $\psi^+\psi$ bosonization.
 With the
help of the standard bosonization trick (here $\tilde J \equiv
\frac{1}{N_c} J^E$)
\be e^{-\Psi\tilde J\Psi}= \int(\det \tilde
J)^{1/2} D\chi\exp [-\chi \tilde J\chi+ i\Psi\tilde J \chi + i\chi
\tilde J \Psi] \label{8} \ee \be Z=\int D\psi D\psi^+ D\chi \exp
L_{QML} \label{9} \ee
 one obtains the effective Quark-Meson Lagrangian (QML)
$$ L^{(2)}_{QML} =\int d^4x\int
d^4y\left\{~^f\psi^+_{a\alpha}(x)[(i\hat\partial+im-i\mu\gamma_4)_{\alpha\beta}\delta(x-y)
+iM^{(fg)}_{\alpha\beta} (x,y)]~^g\psi_{a\beta}(y)- \right.$$ \be
\left.-\frac{1}{N_c}\chi^{(n,k)}(x,y) J^E(x,y)
\chi^{(n,k)}(y,x)\right\} \label{10} \ee
 and the effective  quark-mass operator is
 \be
M^{(fg)}_{\alpha\beta}(x,y) =\sum_{n,k} \chi^{(n,k)}(x,y)
O^{(k)}_{\alpha\beta}t^{(n)}_{fg}\tilde J(x,y).
\label{11}
\ee
Here the operator $\hat{O}$ is a set of all irreducible combinations of Dirac matrixes.

The QML in Eq.(\ref{10}) $L^{(2)}_{QML}$ contains  functions $
\chi^{(n,k)}$  which are integrated out in (\ref{9}), and the
standard way is to find $\chi^{(n,k)}$ from the  stationary point
of $L^{(2)}_{QML}$. Limiting oneself to the scalar and
pseudoscalar fields and using the nonlinear parametrization one
can write for the operator $\hat M$ in (\ref{10})
 \be \hat
M(x,y)=M_{S}(x,y) \hat U(x,y),\hat U=exp(i\gamma_{5}\hat\phi),
\hat \phi(x,y)=\phi^{f}(x,y) t^{f}. \label{12}\ee After
integrating out the quark fields one obtains the ECL in the form
\be L^{(2)}_{ECL}(M_S,\hat \phi)=-2n_f N_c( J^E (x,y))^{-1}
M^2_S(x,y)+ N_c tr\log[(i\hat\partial+im-i\mu\gamma_4)\hat 1+iM_S
\hat U]. \label{13}\ee
 The stationary point equations $\frac{\delta
L^{(2)}_{ECL}}{\delta M_s}= \frac{\delta L^{(2)}_{ECL}}{\delta
\hat \phi}=0$ at $\hat \phi=\hat\phi_0$, $M_s=M_s^{(0)}$
immediately show that $\hat \phi_0=0$ and $M^{(0)}_s$ satisfies
nonlinear equation
\be
 i M^{(0)}_{S}(x,y)={4} tr S J^E(x,y)=
 (\gamma_4 S \gamma_4)  J^E(x,y) ,~~
 S(x,y)=-[i\hat\partial+im -i\mu\gamma_4+iM_S \hat U]^{-1}_{x,y}.
 \label{14}\ee

This equation plays the role of the gap equation and
 is the main point of our further investigation. For $\mu=0$ this was done in \cite{12}
 and in the next section we find how results of \cite{12} are modified by the nonzero $\mu$.

 \section{The confining string at nonzero $\mu$}

Our  basic  equations (\ref{6}),(\ref{7}) are nonlocal in time because of the
integral over $dx_4 dy_4$ in (10). This nonlocality and the parameter
which it governs can be handled most  easily, when one uses instead
of $M(z,z')$, $S(z,z')$ the  Fourier  transforms.

\be
S(z_4-z'_4,\vez,\vez')= \int e^{ip_4(z_4-z'_4)}
S(p_4,\vez, \vez') \frac{dp_4}{2\pi}\label{15}
\ee
and the same for $M(z,z')$. Then from (\ref{14}) one obtains  a
system of equations
\be
(\hat p_4-i\hat \partial_z-im+ i\mu\gamma_4)S(p_4, \vez, \vew) - i\int
M(p_4,\vez, \vez') S(p_4, \vez', \vew) d \vez'=
\delta^{(3)}(\vez - \vew)\label{16}
\ee

To simplify matter, one assumes for $D^E(x)$ the Gaussian form,
$D^E(x)=D(0) \exp \left( -\frac{x^2}{4T^2_g}\right)$. Then  for
$M(p_4, \vez, \vew)$ one has
$$
iM(p_4, \vez, \vew)= 2\sqrt{\pi}T_g \int \frac{dp'_4}{2\pi}
e^{-(p_4-p'_4)^2 T_g^2}\times
 $$
 \be
 \times [J^E(\vez, \vew)\gamma_4S(p'_4,\vez,
 \vew)\gamma_4]\label{17}
 \ee
 where  $J^E$
 is defined in (\ref{7}) and we have factored out the time--dependent
 exponent, using the  Gaussian  representation of $D(u)$.

 All dependence of $M$ on $p_4$ as can be seen in (\ref{17}) is due to the
 factor $\exp [-(p_4-p'_4)^2T_g^2]$    and disappears in the limit when
 $T_g$ goes to zero, while the string tension $\sigma\sim D(0)T^2_g$
 is kept  fixed. This limit can be called the string limit of QCD,
 and we shall study its consequences for equations (\ref{16}),(\ref{17}) in this
 section.

 So in the string limit, with $M$ independent of $p_4$, let us
 consider the Hermitian Hamiltonian
 \be
 \hat H\psi_n \equiv (\frac
 {\alpha_i}{i}\frac{\partial}{\partial z_i}+\beta
 m-\mu)\psi_n(\vez) + \beta \int M (p_4=0, \vez, \vez')\psi_n(\vez') d^3 \vez'= \tilde \varepsilon_n (\mu)\psi_n (\vez)\label{18}
 \ee
with eigenfunctions $\psi_n$ satisfying usual orthonormality
condition
$$
\int \psi^+_n(x)\psi_m( x) d^3 x=\delta_{nm},
$$
From (\ref{18}) it is clear, that one can redefine
 $\tilde\varepsilon_n(\mu) +\mu\equiv\varepsilon_n$, and $\varepsilon_n$ and $\psi_n$ do
not  depend  on $\mu$. Therefore in all subsequent formulas one
can use the same equations as in \cite{12}, but with the
replacement $\varepsilon_n \to \varepsilon_n - \mu$. In
particular, the Green's function $S$ can be expressed as \be
S(p_4,\vex,
\vey)=\sum_n\frac{\gamma_4\psi_n(\vex)\psi^+_n(\vey)}{p_4-i(\varepsilon_n-\mu)}\label{19}
\ee Inserting (\ref{19}) into (\ref{17}) one  has integrals of the
type: \be \int^{\infty}_{-\infty}\frac{dp'_4}{2\pi}\frac{\gamma_4
e^{-(p_4-p'_4)^2T^2_g}} {(p'_4
-i(\varepsilon_n-\mu))}=\frac{i}{2}\gamma_4
sign(\varepsilon_n-\mu)(1+0(p_4T_g,|\varepsilon_n|T_g)\label{20}
\ee

We are  thus led  to the following expression for $M$ in the
string limit \be M(p_4=0, \vez, \vew)= \sqrt{\pi}
T_g J^E(\vez,\vew)\gamma_{4}\Lambda(\vez, \vew)
\label{22}
\ee
where the definition is used
\be
\Lambda(\vez, \vew) = \sum_n\psi_n(\vez) sign (\varepsilon_n-\mu)
\psi^+_n(\vew)\label{23}
\ee

Let us  disregard for the moment the possible appearance in $M$ of
the vector component (proportional to $\gamma_{\mu},\mu=1,2,3,4)$
and concentrate on the scalar contribution only, since that is
responsible for CSB and confinement. Then one can look for
solutions of the Dirac equation (\ref{18}) in the following form
\cite{12} \be \psi_n(\vec r)=\frac{1}{r}\left (
\begin{array}{l}
G_n(r)\Omega_{jlM}\\
iF_n(r)\Omega_{jl'M}
\end{array}
\right)\label{24} \ee where $l'=2j-l$, and introducing the
parameter $\kappa(j,l)=(j+\frac{1}{2}) sign (l-j)$, and replacing
$M$ by a local operator (the generalization to the nonlocal case
is straightforward but cumbersome, for a possible change in the
nonlocal case see \cite{12}), we obtain a system of equations \be
\left\{
\begin{array}{l}
\frac{dG_n}{dr}+\frac{\kappa}{r}G_n-(\varepsilon_n-\mu+m+M_{scal}(r)-M_{vect.}(r))F_n=0\\
\frac{dF_n}{dr}-\frac{\kappa}{r}F_n+(\varepsilon_n-\mu-m-M_{scal}(r)-M_{vect.}(r))G_n=0\\
\end{array}
\right.\label{25} \ee where we suggest $M=M_{scal}+ \gamma_4
M_{vect.}$

Eq.(\ref{24}) possesses a symmetry $(\varepsilon_n-\mu, G_n,
F_n,\kappa)\leftrightarrow (\mu-\varepsilon_n, F_n,G_n,-\kappa)$
which means that for any solution of the form (\ref{23})
corresponding to the eigenvalue $\varepsilon_n-\mu$, there is
another solution of the form \be
\psi_{\mu-\varepsilon_n}(r)=\frac{1}{r}\left (
\begin{array}{l}
F_n(r)\Omega_{jl'M}\\
iG_n(r)\Omega_{jlM}
\end{array}
\right )\label{26}
\ee
corresponding to the eigenvalue $(\mu-\varepsilon_n)$.

Therefore the difference, which enters (\ref{23}) can computed in
terms of $F_n,G_n$ as follows \be \Lambda(\vez, \vew) = \Lambda_0
(\vez, \vew) -\Delta \Lambda (\vez, \vew)\label{26a}\ee where
$\Lambda_0$ is the value of $\Lambda$ for $\mu=0$, i.e.  the same
as in  \cite{12}, while $\Delta \Lambda$ is defined as \be \Delta
\Lambda(\vez, \vew)=  2\sum_{0<\varepsilon_n<\mu}\psi_n(\vez)
\psi^+_n(\vew).\label{27}\ee Using decomposition (\ref{24}) one
can write $\Delta \Lambda$ as \be \Delta \Lambda (\ver, \ver')
=2\sum_{0<\varepsilon_n ,\mu} \left( \begin{array} {ll}
G_nG_n^+ \Omega \Omega^+,&-iG_nF_n^+ \Omega \Omega^{'+}\\
i F_nG_n^+ \Omega' \Omega^+,&F_nF_n^+ \Omega' \Omega^{'+}\end{array}\right) \label{28}\ee
where we have denoted $\Omega\equiv \Omega_{jlM}, \Omega' =\Omega_{jl'M}$,
 and we disregard nondiagonal part of $\Lambda$.

At this point one can follow the relativistic WKB method for Dirac
equation \cite{13} applied
 to calculation of $\Lambda$ in \cite{12} in case of $\mu=0$.
The classically available region for $\psi_n (\ver) $ with energy
$\varepsilon_n \equiv \varepsilon$ is $r_{\min}\leq r \leq
r_{\max}$, where $r_{\max, \min} = \frac{\varepsilon^2 \pm
\sqrt{\varepsilon - 4\sigma^2\kappa^2}}{2\sigma^2}$, and the
summation over $n$ in (\ref{27}) transforms into integration over
$d\varepsilon$, with the lower limit  (for a given $r$)
$\varepsilon_{\min}=\sigma r$. In this way one has for the upper
diagonal element in (\ref{28}). \be \Delta \Lambda (+,+)
=\frac{2\sigma}{\pi^2 r} \delta (1-\cos \theta_{\ver \ver'})
 \int^{\mu/\sigma r}_1 d\tau \frac{\tau+1}{\sqrt{\tau^2-1}}
 \cos (a\sqrt{\tau^2-1}) \theta (\mu-\sigma r)\label{29}\ee
 with $a=\sigma r|r-r'|$, and we keep $r\approx r'$ everywhere except for $a$, since for large
 $a$ (when $r$ is far from $r'$) both $\Lambda$ and  $\Delta \Lambda$ fast decrease.

 In a similar way for the lower diagonal element in (\ref{28}) one has
 \be
 \Delta \Lambda (-,-) =\frac{2\sigma}{\pi^2 r} \delta (1-\cos \theta_{\ver \ver'} )
\int^{\mu/\sigma r}_1 d\tau \frac{\tau-1}{\sqrt{\tau^2-1}}\cos
(a\sqrt{\tau^2-1}) \theta (\mu-\sigma r)\label{30}\ee and taking
$\Lambda_0$ in (\ref{26})  from \cite{12} the resulting form for
$\Lambda$ (\ref{26a}) is
$$
\Lambda( \ver, \ver') \equiv \beta \Lambda_{scalar}+\hat 1 \Lambda_{vector}=
\frac{\beta \sigma}{\pi^2 r}\delta (1-\cos \theta_{\ver \ver'} )
\int_{\mu/\sigma r}^\infty \frac{ d\tau \cos (a\sqrt{\tau^2-1})}{\sqrt{\tau^2-1}} -
$$
\be
-\hat 1 \frac{2\sigma}{\pi^2 r} \delta (1-\cos \theta_{\ver \ver'} )
\int^{\mu/\sigma r}_1  \frac{\tau d\tau}{\sqrt{\tau^2-1}}\cos (a\sqrt{\tau^2-1}).\label{31}\ee

Here $\beta\equiv \gamma_4$, and $\hat 1$ is the unit Dirac matrix, which means,
 that the second term on the  r.h.s. of
(\ref{31}) contributes to the vector part of the resulting mass
operator (\ref{22}),
 while the first term contributes to the scalar part. One
should take into account, that in the first term, \be
\Lambda_{scalar} = \frac{\sigma}{\pi^2 r} \delta (1-\cos
\theta_{\ver \ver'} ) \int^\infty_{\tau_{\min}(\mu)}
\frac{d\tau\cos (a\sqrt{\tau^2-1})}{\sqrt{\tau^2-1}} \label{32}\ee
$\tau_{min}(\mu) =\mu/\sigma r$ for $\mu>\sigma r$ and 1
otherwise, so that for large $r,r \gg \frac{\mu}{\sigma}$, one has
the standard $\mu$-independent value \be \Lambda_{scalar} (r\sim
r'> \mu/\sigma) =\frac{\sigma}{\pi^2 r} K_0 (a) \delta (1-\cos
\theta_{\ver\ver'})\label{33}\ee where we have used relations for the McDonald function $K_0$
 \be
K_0(a) =\int^\infty_0 \frac{\cos ax dx}{\sqrt{1+ x^2}}, ~~
\int^\infty_0 da K_0 (a) =\frac{\pi}{2}.\label{34}\ee

One can check that at large $r, r' (r \sim r' > \mu/\sigma)$
 $\Lambda_{scalar} \approx \Lambda_{scalar}^{ (\mu=0)}$ is a smeared $\delta$ -function,
\be \int \Lambda_{scalar}^{(\mu=0)} (\ver, \ver') d^3\ver'
=1.\label{35}\ee However for large $\mu, \mu\gg \sqrt{\sigma}$, $
\Lambda_{scalar}$ is
  different from $\Lambda_{scalar} (\mu=0)$, and
for $r\sim r' < \mu/\sigma$ one has approximately \be
\Lambda_{scalar} (\ver, \ver') =
\Lambda_{scalar}^{(\mu=0)}-f(r,r') \theta(\mu-\sigma r)
\label{36}\ee where \be f(r, r') = \int^{\mu/\sigma r}_1 \frac{
d\tau \cos (a\sqrt{\tau^2-1})}{\sqrt{\tau^2-1}}=\int^{\lambda_0}_1
\frac{d\lambda}{\sqrt{1+\lambda^2}}\cos a\lambda, \label{37}\ee
 and $\lambda_0 ={ \sqrt{\left(\frac{\mu}{\sigma r}\right)^2-1}}$.

Let us now turn to the vector part of interactions, $\Lambda_{vector}$,
$$
\Lambda_{vector} =- \frac{2\sigma}{\pi^2 r} \delta (1-\cos
\theta_{\ver \ver'} ) \int^{\mu/\sigma r}_1  \frac{\tau
d\tau}{\sqrt{\tau^2-1}}\cos (a\sqrt{\tau^2-1}) \theta (\mu-\sigma
r)=
$$
\be -\hat 1\frac{2\sigma}{\pi^2 r} \delta (1-\cos \theta_{\ver
\ver'} )\frac{\sin (a \sqrt{\left(\frac{\mu}{\sigma
r}\right)^2-1})}{a}. \label{38}\ee Returning back to Eq.
(\ref{22}) one can  deduce, that \be M(p_4 =0, \ver, \ver') =
\sqrt{\pi} T_g J^E(\ver, \ver') [\Lambda_{scal} (\ver, \ver')
+\gamma_4 \Lambda_{vector}]\equiv M_{scal}+ \gamma_4
M_{vect.}\label{39}\ee

Taking into account, that at  $r, r'\gg T_g$ and for the Gaussian
$D^E(x)$ one has from (\ref{7}) \be J^E(\ver,\ver')
\cong\frac{(\ver\ver')}{r r'} 2 T_g \sqrt{\pi} D(0) \min (r,
r')\label{40}\ee one has for $M$ at $r, r'\gg T_g$ and for
 $r\cong r'$ \be M_{scal} (r,r') =\sigma r (\tilde
\delta^{(3)} (\ver, \ver') -\xi (\ver, \ver'))\label{41}\ee where
$\tilde \delta^{(3)} (\ver, \ver') \equiv\Lambda^{(\mu=0)}_{scal}
(\ver, \ver')$ and
 \be \xi (\ver, \ver') =\delta_\mu
=\frac{\sigma}{\pi^2 r} (1-\cos \theta_{\ver\ver'}) f (r,
r')\theta (\mu-\sigma r)\label{42}\ee and $f(r, r')$ is defined in
(\ref{37}).

For $M_{vect}$ one has, using (\ref{38}), \be M_{vect} (\ver,
\ver') =-2\sigma r \varphi_\mu (\ver, \ver') \theta(\mu-\sigma
r)\label{43}\ee
where
 \be \varphi_\mu (\ver,
\ver') =\frac{\sigma}{\pi^2 r} \delta (1-\cos \theta_{\ver \ver'}
)\frac{\sin (a \sqrt{\left(\frac{\mu}{\sigma r}\right)^2-1})}{a}.
\label{44}\ee

Note, that one should actually symmetrize all these expressions,
 e.g. $\frac{1}{r} \to \frac{1}{\sqrt{rr'}}$ etc., but we always are in the regime, where $r\approx r'$.
To estimate the magnitude  of nonlocal kernel $M_{scal} (\ver, \ver')$ and $M_{vect} (\ver, \ver')$ it
is convenient to introduce in (\ref{18}) the local limit of the mass operator, namely
\be \bar M_{scal, vect} (r)=  \int d^3 \ver'
M_{scal, vect}  (\ver, \ver').\label{45}\ee

Exploiting the equalities
\be \int d^3 \ver'\tilde \delta^{(3)}
 (\ver, \ver')=\int d^3 \ver'
\xi  (\ver, \ver')
=\int d^3 \ver'
\varphi_\mu(\ver, \ver')=1\label{46}\ee
one arrives at the expressions
\be
\bar M_{scal}(r) =\sigma r \theta(\sigma r-\mu),~~ \bar M_{vect} = - 2 \sigma r \theta (\mu-\sigma r).\label{47}\ee

In the next section we shall discuss approximations made in deriving (\ref{41}), (\ref{43}), (\ref{47})
and physical implications of these results.

\section{Multiquark states and baryons at nonzero density}

Using the local limit for interactions of a light quark with
string junction, Eq. (\ref{47}), one can calculate the change of
masses of white states due to density.  At this point it is
necessary to define exactly the physical meaning of the parameter
$\mu$, introduced in (\ref{10}) as the quark chemical potential.
That would be true, in the deconfined phase, when quarks can move
freely over the volume  and occupy the Fermi sphere in momentum
space. In the confined phase, however, the term $\mu$ is meant to
be attached to the common string  junction of all quark states,
constituing the complete set of states (\ref{25}). This meaning is
especially clear from Eq.(\ref{22}), where $\mu$ exemplifies the
boundary of the occupied states $\{\varepsilon\}$ in the
$\{n,j,\kappa\}$ notation.

Therefore, assuming that multiquark states located at different
string junction points do not overlap, one can associate $\mu$ for
a given quark with the largest value of $\varepsilon_n$, occupied
by other quarks, belonging to the same string junction point. In
what follows we shall show, that in some situations it will be
advantageous for two or more nucleons (with different string
junctions) to coalesce into a common  state with one  string
junction, building in this way the contracted potential
(\ref{47}).

Note, that for one nucleon (3 quarks) the contraction mechanism
does not work and $\mu$ can be  taken at zero value, since no
forbidden states exist for each quark of a given color. This,
again, is true, provided one-quark states do not overlap at the
given density. The situation changes, however, for two nucleons,
since e.g. one can add to the first nucleon, i.e. 3 quarks in
S-states, 3 quarks in  P-states relative to the same string
junction point. The latter will move in the potential (\ref{47}),
where $\mu$ is equal to the S-state Dirac eigenvalue
$\varepsilon_S$, and the energy $\varepsilon_P
(\mu=\varepsilon_S)$ may become  less than $\varepsilon_S(\mu=0)$:
\be \varepsilon_P(\mu=\varepsilon_S)
<\varepsilon_S(\mu=0)\label{48}\ee The same type of inequality may
occur  for higher $L$ states, when more nucleons coalesce  to the
same Multiquark State (MQS).

This would imply instability of nuclei with respect to transition
to MQS matter.

If, however, Eq.(\ref{48}) does not hold in the limit of zero
density, one may have an interesting situation for increasing
density, large baryonic chemical  potential $\mu_B$. Note, that
$\mu_B\neq 3\mu$, since $\mu$ is attributed to MQS string junction
and does not grow with $\mu_B$ unless different MQS start to
overlap. Hence for $\mu_B>0$ (\ref{48}) should be replaced by
\begin{equation}
\varepsilon_P (\mu =\varepsilon_S)< \varepsilon_S(\mu=0) +
\varepsilon_B (\mathrm{Fermi}). \label{48a}
\end{equation}
To estimate, let's take $\varepsilon_S(\mu=0)\approx 500$~MeV,
$\varepsilon_P(\mu=500~\mathrm{MeV})\approx 630$~MeV (see the
Tables 1,2 below), so for $\varepsilon_B$(Fermi)$\approx 130$~MeV
the phase transition can occur (which corresponds to densities
$3\div 4$ times higher than a nuclear one).

 Below we perform calculations of
$\varepsilon(\mu)$ for different values of $\mu$ and $L=0,1$ We
start with the values of $\varepsilon_{n,L,\kappa} (0)$, which
were computed earlier numerically and analytically using WKB
approximation \cite{12}
(see Table 1).

\begin{table}
\caption{Eigenvalues $\frac{\varepsilon_n}{\sqrt{\sigma}}$
computed numerically, as in (\ref{25})} \label{table1}
\begin{center}
\begin{tabular}{|c|c||c|c|c|}
\hline \multicolumn{2}{|c||}{l} & 0 & 1 & 2 \\
 \hline $n$ & $\mathrm{sign}\kappa$ & \multicolumn{3}{|c|}{ }  \\
\hline 0 & $-$ & 1.29 & 1.93 & 2.40 \\
\hline 0 & $+$ & - & 2.02 & 2.51 \\
\hline 1 & $-$ & 2.34 & 2.76 & 3.11 \\
\hline 1 & $+$ & - & 2.79 & 3.18 \\ \hline
\end{tabular}
\end{center}
\end{table}

%
%
%
%
%
%

In Table 2 we show the change of $\varepsilon_{n L\kappa}$ as
$\Delta M_S=3\Delta\varepsilon_n (S)$ for $S$ and $\Delta M_P (3q)
=2\Delta \varepsilon_n (S)+ \Delta \varepsilon_n (P)$ for the
P-wave baryon  due to nonzero value of $\mu,$ $ \Delta
\varepsilon_{nL\kappa} (\mu) = \varepsilon_{nL\kappa} (\mu)
-\varepsilon_{nL\kappa} (0)$, calculated in perturbation theory,
considering change of potential in (\ref{47}) as a perturbative
parameter. One should note at this point, that the vector part of
interaction written in (\ref{47}) for the case of light quark with
heavy antiquark, and for the case of 3 quarks interacting with
string junction the vector part is transformed as an addition to
the color Coulomb potential with coefficient 1/2 (see \cite{13}
for explicit derivation of scalar and vector interaction in the
nucleon). This prescription was used for calculations in Table 2.

%
%
%

\begin{table}
\caption{Mass  shifts of  $3q$  system  due to nonzero $\mu_q$}
\label{table2}
\begin{center}
\begin{tabular}{|l|l|l|}\hline
$\mu_q$ (MeV)&  $\Delta M _S(3q)$  MeV&$\Delta M_P  (3q)$ MeV\\
\hline 100&-5&-4\\
200&-70&-50\\
300&-266&-200\\
400&-605&-483\\
500&-1023&-876\\\hline
\end{tabular}
\end{center}
\end{table}

\section{ Discussion of results}

Results of the previous section Eqs. (\ref{41}), (\ref{43}), (\ref{47}), can be formulated as follows.
 The relativistic WKB analysis leads to the $\mu$-dependent modification of the confining string, where
  the piece $[0,\mu/\sigma]$ of the string near the
  origin of the string (situated at the  heavy quark position in case of heavy-light quark, or at the string
  junction position in the case of baryons), is dissolved, and the  linear confinement starts beyond the critical
   radius $r_{cr} =\mu/\sigma$. Moreover, an attractive vector interaction appears in the same interval with
    the
   average magnitude $\lan \bar M_{vect}\ran \sim \mu$.

   As shown in Table 2 the change in quark energies
   $\varepsilon_S(\mu)$, $\varepsilon_P(\mu)$ is large and $\Delta
   \varepsilon_P (\mu =400$ MeV $) =- 80 $ MeV, $\Delta
   \varepsilon_P (\mu=500$ MeV)       $=-190$ MeV, while
   $|\varepsilon_P(\mu=0) - \varepsilon_S(\mu=0) |\approx 320$
   MeV. Hence it seems to be not advantageous for a quark in
   neighboring nucleon to add in the P-wave to the 3q system in $S$
   state, with a reduced confinement.

   However, the gap is not large. Indeed, the MQS with five quarks
   in $S$ state and one quark in $P$ state will have the mass shift $\Delta$ only 130 MeV above the NN mass.
   Note, that  phenomenological
   value of the $6q$ MQS $\Delta\cong 210$ MeV is in the same
   ballpark \cite{15a}. Hence for higher density this value of 130
   MeV can  be compensated by baryon Fermi energy, and the  baryon
   matter can become unstable for transition into MQS matter.

   These conclusions should be taken as qualitative.
First of all, the WKB method is not a good approximation  at small distances, and  we have omitted
 exponentially damped part of $\psi_n(\vez)$ in  the spectrum,
  therefore the inner part of the string is to some extent delocalized (see \cite{12}
 for details) and smoothed, as shown in Fig.1.

\begin{figure}
\caption{Interactions $\bar M_{scal} (r)$ (solid line) and $\bar
M_{vect}(r)$  (dashed line)as functions  of distance $r$, with $b=
\mu/\sigma$. Dotted line and dashed line show the qualitative
smoothed form of both terms respectively.} \label{fig1}
\includegraphics[width=120mm,keepaspectratio=true]{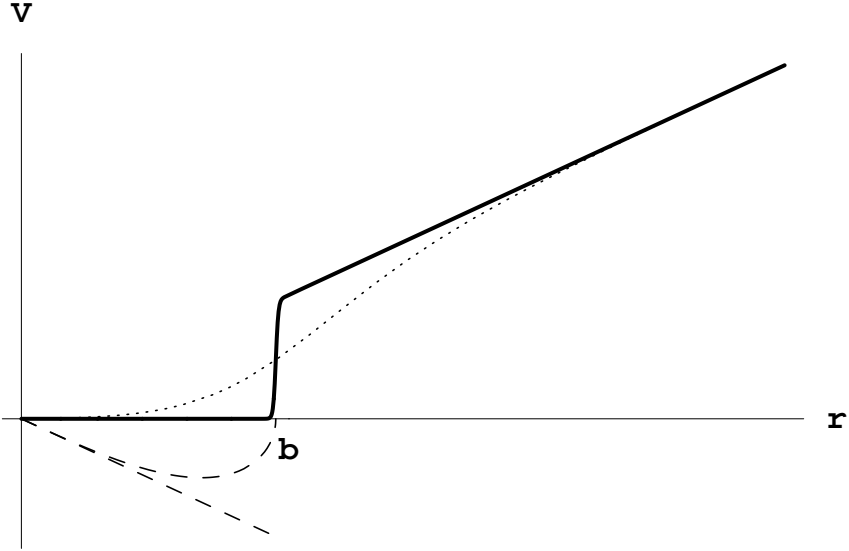}
\end{figure}


Secondly, the interaction between MQS and baryons was not
considered above.

Thirdly,  we have not taken into account a possible modification
and destruction of the vacuum due to  the influence of high
density quark matter, which might decrease $\sigma$ or cancel the
string completely (as it is happens
 in the thermal phase transition \cite{6}).
 We have calculated in \cite{5} the critical value of
 $\mu_q(T=0)$, where deconfinement due to density occurs
 $\mu_q^{crit}=\frac{V_1(T^{(0)})}{2} + \left( \frac{9\pi^2}{16
 n_f}G_2\right)^{1/4}\cong 0.6$ GeV, which corresponds to 7 times
 normal nuclear density. We are interested here however in the
 confinement region and small temperatures, with density few times
 larger than normal.

If however, no density
induced vacuum deconstruction
takes place, then the resulting physical picture according to Eqs. (\ref{47}), is
the net decreasing of confinement in the inner region of some ensemble of quarks, and appearance
of attractive vector potential of the order of $\mu$ acting on each quark.
 This may cause creation of partly  deconfined
bubbles consisting of $3n$ quarks, $n=2,3,..$ in the midst of the
nuclear medium, and dynamically is similar to the $3n q$ bag
formation, which  was studied  before in the framework of the
Quark Compound Bag model \cite{15a}.  Note, however, that bag
boundary conditions might be strongly
 modified as compared to the standard MIT bag model. As it is,
 preliminary estimates in the previous section demonstrate a
 possible formation of a new phase of MQS at rather high densities
 (in heavy ion collisions or in neutron stars) when neutron matter
 density exceeds 3-4 normal values.
A quantitative analysis of this situation needs a more accurate
consideration of
 the $3n$ quark system using nonlinear equations for the $3nq$
 Green's function, generalizing Eq. (\ref{14}).

 The formation of these high-density $3nq$ bubbles may be connected with the explanation
 so-called cumulative effects in the hadron-nucleus(and nucleus-nucleus) collisions,\cite{15b}
  for an example of this discussion
see \cite{16} and refs. therein.

The authors are grateful to N.O.Agasian and A.B.Kaidalov for
useful discussions.

 This work was supported by the Russian State Corp. ``Rosatom'',
by the President Grant No. NSh-4961.2008.2 for the leading
scientific schools, and by the grants RFFI-08-02-00657 and
RFFI-09-02-00629. One of the authors (M. A. T.) would like to
acknowledge the partial support from the President Grant No.
MK-2130.2008.2.

\end{document}